\documentclass[sigconf]{acmart}

\AtBeginDocument{%
  \providecommand\BibTeX{{%
    \normalfont B\kern-0.5em{\scshape i\kern-0.25em b}\kern-0.8em\TeX}}}

\setcopyright{acmcopyright}
\copyrightyear{2025}
\acmYear{2025}
\setcopyright{rightsretained}
\acmConference[ETRA '25]{2025 Symposium on Eye Tracking Research and Applications}{May 26--29, 2025}{Tokyo, Japan}
\acmBooktitle{2025 Symposium on Eye Tracking Research and Applications (ETRA '25), May 26--29, 2025, Tokyo, Japan}\acmDOI{10.1145/3715669.3726801}
\acmISBN{979-8-4007-1487-0/2025/05}

\begin{document}

\title{eye2vec: Learning Distributed Representations of Eye Movement for Program Comprehension Analysis}

\author{Haruhiko Yoshioka}
\affiliation{%
  \institution{Nara Institute of Science and Technology}
  \city{Ikoma}
  \state{Nara}
  \country{Japan}
}
\email{yoshioka.haruhiko.yi4@naist.ac.jp}

\author{Kazumasa Shimari}
\affiliation{%
  \institution{Nara Institute of Science and Technology}
  \city{Ikoma}
  \state{Nara}
  \country{Japan}}
\email{k.shimari@is.naist.jp}

\author{Hidetake Uwano}
\affiliation{%
  \institution{National Institute of Technology, Nara College}
  \city{Yamato-koriyama}
  \state{Nara}
  \country{Japan}
}
\email{uwano@info.nara-k.ac.jp}

\author{Kenichi Matsumoto}
\affiliation{%
 \institution{Nara Institute of Science and Technology}
 \city{Ikoma}
 \state{Nara}
 \country{Japan}}
\email{matumoto@is.naist.jp}


\begin{abstract}
This paper presents \textit{eye2vec}, an infrastructure for analyzing software developers’ eye movements while reading source code.
In common eye-tracking studies in program comprehension, researchers must preselect analysis targets such as control flow or syntactic elements, and then develop analysis methods to extract appropriate metrics from the fixation for source code.
Here, researchers can define various levels of AOIs like words, lines, or code blocks, and the difference leads to different results.
Moreover, the interpretation of fixation for word/line can vary across the purposes of the analyses.
Hence, the eye-tracking analysis is a difficult task that depends on the time-consuming manual work of the researchers.
\textit{eye2vec} represents continuous two fixations as transitions between syntactic elements using distributed representations.
The distributed representation facilitates the adoption of diverse data analysis methods with rich semantic interpretations.
\end{abstract}



\maketitle

\section{Introduction}
\label{intro}
Analyzing developers' eye movements is crucial for understanding their program comprehension processes.
For example, in a method summarization task, analyzing the developers' eye movements while reading source code has revealed that experts tend to generate summaries based on the most frequently read lines~\cite{Abid2019}.
Extracting the understanding patterns and strategies of proficient developers from their eye movements can be beneficial for enhancing development efficiency and educating developers.

When analyzing eye movements, researchers need to determine which syntactic elements or methods to focus on based on the fixation data mapped onto the source code.
This choice largely depends on the researchers' experience, which is a major reason for the high difficulty of conducting eye-tracking analyses.
Additionally, analyzing multiple meanings simultaneously is challenging, such as combining an analysis that focuses on specific syntactic elements of source code with one that examines the structural significance of eye movement patterns.
Therefore, many existing studies have focused on only a single meaning~\cite{Abid2019, Sharafi2022, Lin2016}.
To overcome these challenges, it is desirable to develop an analytical method that automatically captures distinctive eye movement features and considers multiple meanings.

\textit{eye2vec} is a novel solution that combines developers' eye movement data with distributed representations of source code. \textit{eye2vec} supports semantic-based analysis and addresses the issues of analysis being dependent on researchers' experience.
Rather than mapping to display coordinates or lines, \textit{eye2vec} mechanically maps eye movement data to syntactic elements, thereby enabling the representation of eye movement in forms that can be interpreted as words, sentences, blocks, and other semantic units.
Moreover, distributed representations can capture multiple semantic aspects, and by leveraging machine learning and deep learning, it becomes possible to conduct multifaceted analyses in an automated and comprehensive manner using eye movement data.

\section{eye2vec: Learning Distributed Representations of Eye Movement}
\label{methods}
\begin{figure*}[t]
    \centering
    \includegraphics[width=1.0\linewidth]{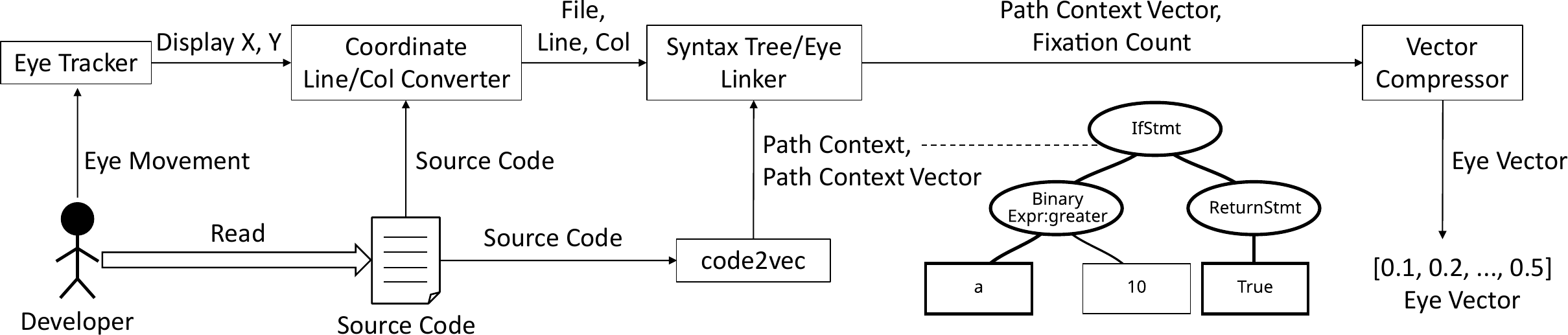}
    \Description{Overview of the eye2vec architecture}
    \caption{Overview of the eye2vec architecture}
    \label{fig:eye2vec}
\end{figure*}

\textit{eye2vec} converts coordinate-based eye movements recorded by the eye tracker into distributed representations.

Figure~\ref{fig:eye2vec} provides an overview of \textit{eye2vec}.
Rectangles represent modules, and thin arrows indicate the flow of information.
At first, a developer reads the source code, and the eye tracker records the developer's eye movement.
Coordinate Line/Column Converter converts the coordinate-based eye movements into line and column numbers in the source code.
In this study, this process is performed using the tool \textit{iTrace}~\cite{Guarnera2018}.
In parallel, \textit{code2vec}~\cite{alon2019code2vec} extracts path contexts, which refer to the path between two words on the abstract syntax tree (AST) generated from source code.
By collecting the paths that the AST follows, it becomes possible to perform eye movement analysis that takes into account multiple meanings.
For example, \textit{eye2vec} can interpret eye movements from variable A in the if block within a loop statement to variable B outside the loop statement.
\textit{eye2vec} can take into account the complex structure of source code, such as loops and branches.
Then, \textit{code2vec} outputs these path contexts along with their corresponding learned vectors, which represent the features of the source code.
Syntax Tree/Eye Linker associates the line and column numbers with the path contexts and converts the eye movements into vectors and fixation counts of path contexts.
Finally, Vector Compressor aggregates vectors and fixation counts into an eye vector.
This eye vector represents the eye movement feature in program comprehension.

\textit{code2vec} is used to convert the eye movement from one word to another word into a vector.
\textit{code2vec} is one of the methods for obtaining distributed representations, and it converts the relationship between two words in source code (i.e., path context) into a pre-trained embedding vector.
By using path contexts, the sequence of fixations while reading source code can be represented as a series of eye movements between two semantically related words, such as a method's caller/callee or a variable's declaration/usage.
\textit{eye2vec} assigns weights to important path contexts that reflect the developer's characteristics.
Specifically, Syntax Tree/Eye Linker converts the number of fixations on each path context into a ratio, and Vector Compressor applies weights to the embedding vectors accordingly.
In this way, the more frequently a developer focuses on a particular path context, the more its corresponding vector is emphasized, reflecting the distinctive characteristics of the developer's eye movement.
Finally, by aggregating multiple vectors generated from each path, \textit{eye2vec} generates an eye vector representing the semantics of source code.

The generated eye vector can be applied to automated analyses such as deep learning because it can capture higher-order semantic representations. It leads to interpreting eye movements in the context of comprehension strategy.
Furthermore, a generated eye vector is also a distributed representation integrating the relations of each path context that captures the relationship between two words. This structure-based information makes it possible to perform low-level analyses, such as tracking the variable $a$ operations.

\section{USE CASES}
\label{use cases}

\subsection{Support for Data-Mining Analyses}
The distributed representations generated by \textit{eye2vec} embed rich semantic information and express meaning as machine-interpretable numerical data, in comparison to conventional analyses based on coordinates or line numbers.
It enables researchers to conduct comprehensive and automated analyses by applying various data processing techniques.
As a result, researchers can conduct exploratory data-mining analyses in addition to hypothesis-based statistical analysis.
For example, researchers can investigate the underlying significance of trends observed in groups of semantically similar syntactic elements.

\subsection{Label Prediction Using Machine Learning}
By applying \textit{eye2vec}, label prediction becomes feasible based on developers' eye movement data and labels representing their characteristics.
We can predict factors such as developers’ proficiency in specific languages and their level of program comprehension.
Consequently, educators can provide adaptive instructional support tailored to the unique characteristics of individual developers.

\section{Conclusions and Future Work}
\label{conclusions}
This study proposes \textit{eye2vec}, an infrastructure for analyzing developers' eye movements based on semantic information.
It enables automated analyses that consider multiple semantic aspects.
\textit{eye2vec} also allows researchers to measure the similarity between comprehension patterns as distances in a vector space.

\textit{eye2vec} extracts comprehension patterns, so it must not be linked to any personal information. Therefore, the current version of \textit{eye2vec} only accepts eye movement data as input.

Future research directions include developing utilities that apply machine learning and deep learning techniques to the distributed representations generated by \textit{eye2vec}.
The goal is to support other researchers in efficiently extracting useful comprehension patterns from eye movement data through automated semantic analysis.

\begin{acks}
This work has been supported by JSPS KAKENHI Nos. JP20H05706, JP21K11842, and JP23K16862.
\end{acks}


\bibliographystyle{ACM-Reference-Format}
\bibliography{references}

\end{document}